\documentclass[journal,comsoc,onecolumn, twoside]{IEEEtran}
\usepackage{cite}
\usepackage{color}
\usepackage{amsmath,amssymb}
\usepackage{amscd,amsfonts}
\usepackage{graphicx}
\usepackage{epsf}
\usepackage{epsfig}
\usepackage{multirow}
\usepackage{array}
\usepackage{placeins}
\usepackage{amsthm}
\usepackage{subfigure}
\usepackage{upgreek}
\usepackage[T1]{fontenc}
\usepackage[pdfborder={0 0 0},colorlinks=false]{hyperref}
\usepackage{scalerel}
\usepackage{tikz}

\usetikzlibrary{svg.path}
\definecolor{orcidlogocol}{HTML}{A6CE39}
\tikzset{
  orcidlogo/.pic={
    \fill[orcidlogocol] svg{M256,128c0,70.7-57.3,128-128,128C57.3,256,0,198.7,0,128C0,57.3,57.3,0,128,0C198.7,0,256,57.3,256,128z};
    \fill[white] svg{M86.3,186.2H70.9V79.1h15.4v48.4V186.2z}
                 svg{M108.9,79.1h41.6c39.6,0,57,28.3,57,53.6c0,27.5-21.5,53.6-56.8,53.6h-41.8V79.1z M124.3,172.4h24.5c34.9,0,42.9-26.5,42.9-39.7c0-21.5-13.7-39.7-43.7-39.7h-23.7V172.4z}
                 svg{M88.7,56.8c0,5.5-4.5,10.1-10.1,10.1c-5.6,0-10.1-4.6-10.1-10.1c0-5.6,4.5-10.1,10.1-10.1C84.2,46.7,88.7,51.3,88.7,56.8z};
  }
}
\newcommand\orcidicon[1]{\href{https://orcid.org/#1}{\mbox{\scalerel*{
\begin{tikzpicture}[yscale=-1,transform shape]
\pic{orcidlogo};
\end{tikzpicture}
}{|}}}}

\begin{document}

\title{{Design of QAM-FBMC Waveforms Considering MMSE Receiver}}

\author{Hyungsik~Han$^{\orcidicon{0000-0002-2196-9377}}$,~\IEEEmembership{Member,~IEEE,}
        Namshik Kim,
       and~Hyuncheol~Park,~\IEEEmembership{Senior Member,~IEEE}


\thanks{
H. Han, N. Kim and H. Park are with School of Electrical Engineering, Korea Advanced Institute of Science and Technology (KAIST), Daejeon,
 Korea 34141 (e-mail: \{crezol, nskim73, hcpark\}@kaist.ac.kr). }
\thanks{Digital Object Identifier 10.1109/LCOMM.2019.2952375}
}

\markboth{Accepted to IEEE COMMUNICATIONS LETTERS, 2019}%
{HAN, KIM, AND PARK: \MakeUppercase{Design of QAM-FBMC Waveforms Considering MMSE Receiver}}

%



\maketitle

\begin{abstract}
Due to its high spectral confinement characteristics and spectral efficiency, QAM-FBMC is considered a candidate waveform to replace CP-OFDM. {\color{black}QAM-FBMC has inevitable non-orthogonality both in time and frequency, and the system and filter must be well-designed to minimize the interferences.} However, existing QAM-FBMC studies utilize a matched filter as the receiver filter, which is not suitable for a non-orthogonal system. Therefore, in this paper, we design the prototype filters considering the MMSE criterion, {\color{black}and propose a system providing the highest SINR in QAM-FBMC which cannot avoid non-orthogonality.} In addition, we confirm that the proposed filters show best performance at target SNR than the reference filters.

\end{abstract}


%
\IEEEpeerreviewmaketitle

\section{Introduction}\label{sec:intro}

Filter-bank multi-carrier (FBMC) has been considered an alternative waveform to resolve the disadvantages of cyclic prefix-orthogonal frequency division multiplexing (CP-OFDM). {\color{black}The advantages of FBMC includes such as high out-band emission characteristics and the loss of spectral efficiency \cite{banelli2014modulation}.} By subband filtering in the up-sampled frequency domain, the FBMC improves the spectral confinement characteristics dramatically, and can increase the spectral efficiency by reducing the guard band and the cyclic prefix. However, the conventional FBMC utilizes the offset-QAM (OQAM) for maintaining the orthogonality and this causes complicated signal processing on complex channels (such as multiple-input and multiple-output (MIMO) operation) due to the intrinsic interference problem \cite{bellanger2010fbmc,zakaria2012novel}.

To avoid the problem of OQAM-FBMC, QAM-FBMC using non-orthogonal prototype filters has been proposed \cite{yun2015new}. Instead of maintaining the orthogonality in the real domain through OQAM, {\color{black}QAM-FBMC uses QAM which increases the risk of giving up some degree of orthogonality in the complex domain.} A QAM-FBMC filter designed to be non-orthogonal may degrade performance, but signal processing in the complex domain can easily be applied to QAM-FBMC in a way similar to CP-OFDM. The remaining major issue of QAM-FBMC is the need to design the system and filters to minimize the non-orthogonality that causes performance degradation.

There have been several previous studies on the problem of designing filters for QAM-FBMC. In the initial study, it was proposed to maximize the self-signal-to-interference ratio (self-SIR) by utilizing multiple base filters \cite{yun2015new,kim2016introQAMFBMC}. In these studies, the QAM-FBMC system utilized different prototype filters for even and odd subcarriers, making the system complex. These filters were also poorly localized in the time domain, which presents a vulnerability problem for multipath channels in the FBMC system without CP. To overcome this problem, some studies have proposed design of a single prototype filter considering the localization\cite{kim2017waveform,han2019design}. Compared to the initial studies, a system with the prototype filter becomes simpler and stronger than the selective channel, but shows a slight decrease in self-SIR.

Although a variety of filter-design studies have been carried out, these previous studies have commonly utilized a matched filter as the receiver filter. Unlike the systems that achieve orthogonality through a matched filter (e.g., OFDM or OQAM-FBMC), \color{black} QAM-FBMC cannot be orthogonal.
Therefore, for QAM-FBMC, a matched filter is not suitable as a receiver filter.
\color{black} Instead, a filter following the minimum mean square error (MMSE) criterion can minimize interference and maximizes SINR, and be a more suitable linear receiver for QAM-FBMC \cite{poor2013introduction}.
In addition, by designing the prototype filter considering the MMSE receiver filter, we can achieve the system close to orthogonal waveforms as possible, and also expect to significantly mitigate the BER performance degradation due to the non-orthogonal filters.
\color{black}

In the work reported in this paper, we design a prototype filter considering the MMSE receiver filter, and compare its performance with that of an existing filter using a matched receiver filter.
\color{black} Specifically, in Section \ref{sec:sysmodel}, we describe a simplified system model of a QAM-FBMC transceiver as a stacked matrix representation. This simplified system model makes signal processing easier compared to the matrix sum model that follows the traditional overlap-and-sum structure.
In Section \ref{sec:mmse}, we formulate a receiver filter matrix that follows the MMSE criterion using the simplified system model.
In Section \ref{sec:designwaveform}, we update the conventional filter design problem to apply the simplified system model, and present the prototype filter coefficients obtained by performing global optimization. \color{black} 
In Section \ref{sec:simul}, the simulation results show that the proposed prototype filters with a MMSE receiver filter have better performance than the reference filters that utilize the matched filter as the receiver filter. Finally, conclusions are drawn in Section \ref{sec:conclusion}.

\begin{figure}[t!]
\centering
\includegraphics[draft=false,width=1.0\columnwidth]{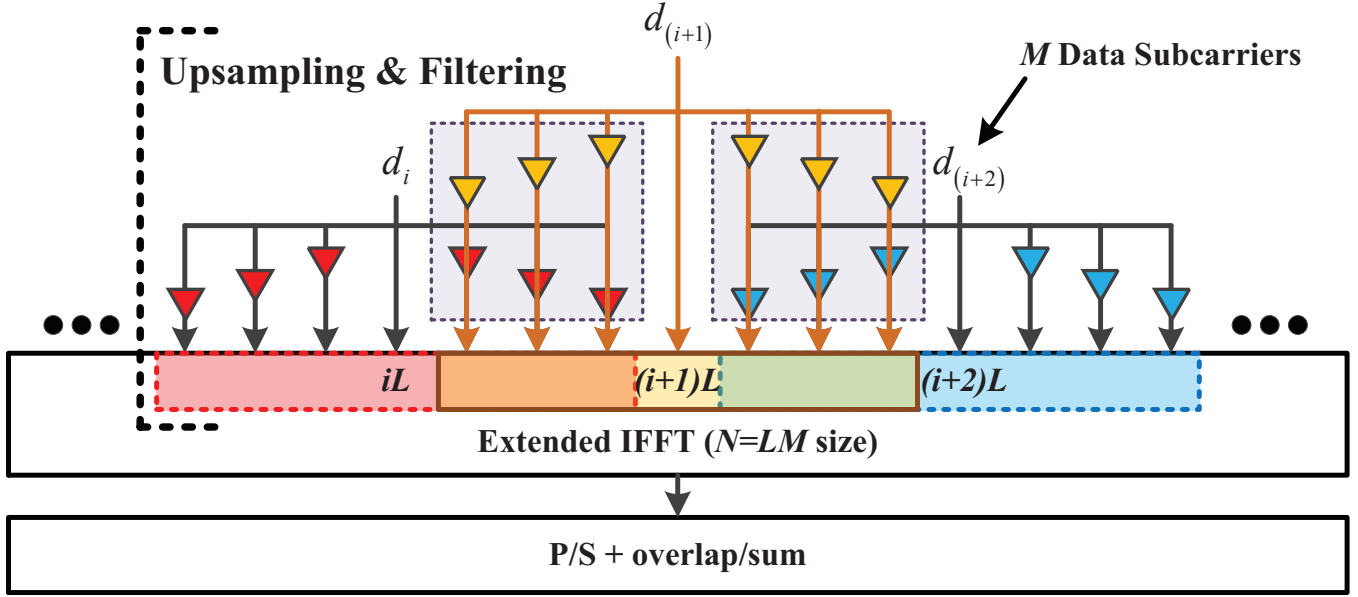}
\caption{The QAM-FBMC transmitted signal structure ($L=4$).\label{fig:fbmc_struct}}
\vspace{-2mm}
\end{figure}


\section{System Model of QAM-FBMC Transceiver}\label{sec:sysmodel}

In this section, we write the system model of QAM-FBMC to formulate the MMSE criterion of the receiver. As can be seen in \cite{yun2015new} and \cite{kim2017waveform}, the system model of QAM-FBMC can be written in matrix form as the sum of the overlapped signals. \color{black} With this existing system model, it is difficult to represent and formulate MMSE criterion. However, in this paper, we change the model to a matrix representation with stacked form of the data vectors contained in the overlap-and-sum structure. At the end of this section, we will represent the QAM-FBMC system model in the form of a simple linear system. \color{black}

We denote $M$ as a number of subcarriers, $L$ as an overlapping factor for the filtering in the frequency domain, and $N=LM$ as a number of upsampled frequency points, as shown in Fig. \ref{fig:fbmc_struct}.

\subsection{Transmitted Signal Model}

To formulate the stacked representation of overlapped transmit signal, we define the $k$-th stacked data symbol vector $\overline {\bf{d}} \left[ k \right]$ as follows:
\begin{equation}\label{eq:stack_datasymbol}
\overline {\bf{d}} \left[ k \right]  = {\left[ {{\bf{d}}{{\left[ {k - L} \right]}^T}, \cdots ,{\bf{d}}{{\left[ k \right]}^T}, \cdots ,{\bf{d}}{{\left[ {\left( {k + L - 1} \right)} \right]}^T}} \right]^T},
\end{equation}
where ${{\bf{d}}\left[ k \right]}$ is a $k$-th data vector with the $m$-th element, ${d_m}\left[ k \right]$, which is a QAM data symbol. The $\overline {\bf{d}} \left[ k \right]$ includes the $\left( {k - L + 1} \right)$-th  to the $\left( {k + L - 1} \right)$-th data symbols considering the overlap-and-sum structure, and includes $\left(k - L\right)$-th symbol to consider interference from channel delays.

From the $\overline {\bf{d}} \left[ k \right]$, a $k$-th transmitted signal ${\bf{\bar x}}\left[ k \right]$ in the stacked representation can be written as
\begin{equation}\label{eq:stacked_x_k}
{\bf{\bar x}}\left[ k \right] = {\overline {\bf{G}} _f}\overline {\bf{d}} \left[ k \right].
\end{equation}
And ${\overline {\bf{G}} _f}$ is a stacked pulse-shaping filter matrix of size $\left( {M+N} \right) \times 2N$ as
\begin{equation}
{\overline {\bf{G}} _f} = \left[ {{\bf{G}}_f^{\left( { - L} \right)}, \cdots ,{\bf{G}}_f^{\left( 0 \right)}, \cdots ,{\bf{G}}_f^{\left( {L - 1} \right)}} \right].
\end{equation}
The ${\bf{G}}_f^{\left( l \right)}$ is constructed by $m$-th column ${\bf{g}}_m^{\left( l \right)}$ with $\left( n+M \right)$-th element $g_m^{\left( l \right)}\left[ n \right]$, and $g_m^{\left( l \right)}\left[ n \right]$ is defined in $n =  - M, \cdots ,N - 1$ range as follows:
\begin{equation}
\begin{aligned}
g_m^{\left( {l \ge 0} \right)}\left[ n \right] &= \left\{ \!\!
    {\begin{array}{*{20}{l}}
        {p_0\left[ n \right]{e^{\left( {j2\pi \frac{{m\left( {n - lM} \right)}}{M}} \right)}}},&{n = lM, \cdots ,N - 1,}\\
        0,&{{\rm{otherwise,}}}
    \end{array}} \right.\\
g_m^{\left( {l < 0} \right)}\left[ n \right] &= \left\{ \!\!
    {\begin{array}{*{20}{l}}
        {p_0\left[ n \right]{e^{\left( {j2\pi \frac{{m\left( {n - lM} \right)}}{M}} \right)}}},&{n =  - M, \cdots ,N + lM - 1,}\\
        0,&{{\rm{otherwise,}}}
    \end{array}} \right.
\end{aligned}
\end{equation}
where $p_0\left[ n \right]$ is a prototype filter defined in $n = 0, \cdots, N-1$ range. By equation \eqref{eq:stacked_x_k}, we can now represent the transmitted signal that has passed the overlap-and-sum structure without sum operation.

\subsection{Received Signal Model}\label{sec:rx_model}

To represent the time-domain channel for the stacked vector ${\bf{\bar x}}\left[ k \right]$, we define $\left( {N + M} \right) \times \left( {N + M} \right)$ channel convolution matrix ${\bf{H}}$, and the each column of the matrix is given by shift of the channel impulse response with $L_c$ taps as follows:
\begin{equation}
{\left[ {\bf{H}} \right]_{\left( {{\rm{:,}}} m \right)}} = {\rm{shift}}\left\{ {{{\left[ {{h_0}\, \cdots \,{h_{{L_c} - 1}}\,\,{{\bf{0}}_{N + M - {L_c}}}} \right]}^T},\,m - 1} \right\}.
\end{equation}

\begin{figure}[t!]
\centering
\includegraphics[draft=false,width=1.0\columnwidth]{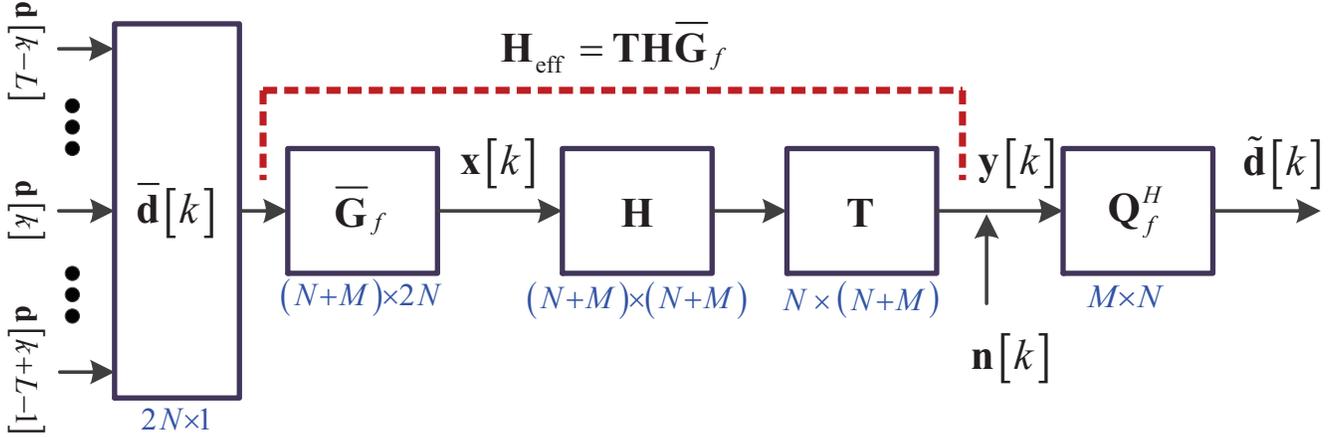}\vspace{-2mm}
\caption{The stacked matrix representation of the QAM-FBMC system.\label{fig:trx_struct}}
\vspace{-3mm}
\end{figure}

The $k$-th received signal vector ${\bf{y}}\left[ k \right]$ of size $N \times 1$ can be written as
\begin{equation}\label{eq:stacked_rx}
    {\bf{y}}\left[ k \right] = {\bf{TH}}{\overline {\bf{G}} _f}\overline {\bf{d}} \left[ k \right] + {\bf{n}}\left[ k \right],
\end{equation}
where ${\bf{T}} = \left[ {{{\bf{0}}_{N \times M}}{\rm{\;\;}}{{\bf{I}}_N}} \right]$ is a time-domain slice matrix to extract the samples in the $k$-th received window, and ${\bf{n}}\left[ k \right]$ is the additive white Gaussian noise (AWGN) vector with complex Gaussian distribution of $CN\left( {{\bf{0}},\sigma _n^2{{\bf{I}}_N}} \right)$.
If we simply consider linear receiver process as a filter of size $M \times N$, we can represent the $k$-th detected data symbol as follows:
\begin{equation}
{\bf{\tilde d}}\left[ k \right] = {\bf{Q}}_f^H{{\bf{H}}_{\rm{eff}}}\overline {\bf{d}} \left[ k \right] + {\bf{Q}}_f^H{\bf{n}}\left[ k \right],
\end{equation}
where ${{\bf{H}}_{\rm{eff}}} = {\bf{TH}}{\overline {\bf{G}} _f}$, and ${{\bf{Q}}_f}$ is a receiver filter matrix that we will formulate by the MMSE criterion in section \ref{sec:mmse}. Overall, we show the QAM-FBMC transceiver structure in stacked matrix representation in Fig. \ref{fig:trx_struct}.

\section{Receiver Filter by MMSE Criterion}\label{sec:mmse}

In this section, we formulate the receiver filter matrix ${{\bf{Q}}_f}$ that follows the MMSE criterion. To do this, we set a MMSE problem of the $k$-th detected data symbol as follows:
\begin{equation}\label{eq:MMSE_problem}
\begin{aligned}
{\bf{\hat Q}}_f^{\left( {{\rm{MMSE}}} \right)}
&=\mathop {\arg \min }\limits_{{{\bf{Q}}_f}} E\left[ {\left\| {\bf{e}} \right\|_2^2} \right]
= \mathop {\arg \min }\limits_{{{\bf{Q}}_f}} E\left[ {\left\| {{{\bf{\tilde d}}\left[ k \right]} - {\bf{S}}\overline {\bf{d}} \left[ k \right]} \right\|_2^2} \right]\\
 &= \mathop {\arg \min }\limits_{{{\bf{Q}}_f}} E\left[ {\left\| {\left( {{\bf{Q}}_f^H{{\bf{H}}_{\rm{eff}}}\overline {\bf{d}} \left[ k \right] + {\bf{Q}}_f^H{\bf{n}}} \right) - {\bf{S}}\overline {\bf{d}} \left[ k \right]} \right\|_2^2} \right],
\end{aligned}
\end{equation}
where ${\bf{S}} = \left[ {{{\bf{0}}_{M \times N}}, \cdots ,{{\bf{I}}_M}, \cdots {{\bf{0}}_{M \times \left( {L - 1} \right)M}}} \right]$ is a matrix for extracting the $k$-th data symbol from the stacked vector ${\overline {\bf{d}} \left[ k \right]}$. The expectation of the squared-error becomes
\begin{equation}\label{eq:expect_err}
\begin{aligned}
E\left\{ {\left\| {\bf{e}} \right\|_2^2} \right\} &= {\rm{tr}}\left\{ {\left( {{\bf{Q}}_f^H{{\bf{H}}_{{\rm{eff}}}} - {\bf{S}}} \right)\left( {{\bf{H}}_{{\rm{eff}}}^H{{\bf{Q}}_f} - {{\bf{S}}^H}} \right)} \right\}\sigma _d^2\\
 &+ {\rm{tr}}\left\{ {{\bf{Q}}_f^H{{\bf{Q}}_f}} \right\}\sigma _n^2,
\end{aligned}
\end{equation}
where we assume that the covariance of the stacked data symbol vector is $E\left[ {\overline {\bf{d}} {{\overline {\bf{d}} }^H}} \right] = \sigma _d^2{{\bf{I}}_{2N}}$, and the data symbols and AWGN are uncorrelated. We can rewrite the \eqref{eq:expect_err} as follows:
\begin{equation}\label{eq:expect_err2}
\begin{aligned}
&E\left\{ {\left\| {\bf{e}} \right\|_2^2} \right\} = {\rm{tr}}\left\{ {{\bf{Q}}_f^H{\bf{A}}{{\bf{Q}}_f} - {\bf{Q}}_f^H{\bf{B}} - {{\bf{B}}^H}{{\bf{Q}}_f} + \sigma _d^2{\bf{I}}} \right\}\\
 &= {\rm{tr}}\left\{ {{{\left( {{{\bf{C}}^H}{{\bf{Q}}_f} - {{\bf{C}}^{ - 1}}{\bf{B}}} \right)}^H}\left( {{{\bf{C}}^H}{{\bf{Q}}_f} - {{\bf{C}}^{ - 1}}{\bf{B}}} \right) + {{\bf{e}} _{\min }}} \right\},
\end{aligned}
\end{equation}
where ${\bf{A}} = \sigma _d^2{{\bf{H}}_{{\rm{eff}}}}{\bf{H}}_{{\rm{eff}}}^H + \sigma _n^2{{\bf{I}}_N}$, ${\bf{B}} = {{\bf{H}}_{{\rm{eff}}}}{{\bf{S}}^H}\sigma _d^2$, ${\bf{A}} = {\bf{C}}{{\bf{C}}^H}$ (can be factorized by singular value decomposition), and ${{\bf{e}} _{\min }} = \sigma _d^2{{\bf{I}}_{2N}} - {{\bf{B}}^H}{{\bf{A}}^{ - 1}}{\bf{B}}$.
\color{black} Since ${{\bf{e}} _{\min }}$ is not related to ${{{\bf{Q}}_f}}$, and the left term is the form of Frobenius norm, the equation \eqref{eq:expect_err2} becomes convex \cite{boyd2004convex}. Therefore, the minimum point is ${{\bf{C}}^H}{{\bf{Q}}_f} - {{\bf{C}}^{ - 1}}{\bf{B}} = 0$, and the solution of the MMSE problem becomes \color{black}
\begin{equation}\label{eq:Qf_MMSE}
\begin{aligned}
{\bf{\hat Q}}_f^{\left( {{\rm{MMSE}}} \right)}
= {{\bf{C}}^{ - H}}{{\bf{C}}^{ - 1}}{\bf{B}} = {{\bf{A}}^{ - 1}}{\bf{B}}
= {\left( {{{\bf{H}}_{{\rm{eff}}}}{\bf{H}}_{{\rm{eff}}}^H + \frac{{\sigma _n^2}}{{\sigma _d^2}}{\bf{I}}} \right)^{ - 1}}\!\!{{\bf{H}}_{{\rm{eff}}}}{{\bf{S}}^H}.
\end{aligned}
\end{equation}

\section{Waveform Design with MMSE Receiver}\label{sec:designwaveform}

We define the QAM-FBMC waveform design problem as the optimization of the prototype filter. We assume that the single prototype filter for practical transmission structure, and prototype filter is complex modulated by the frequency coefficients as follows:
\begin{equation}\label{eq:proto_timefunction}
{p_0}\left( n \right) = \sum\limits_{k =  - \left( {K - 1} \right)}^{K - 1} {{q_k}{e^{j2\pi \frac{{kn}}{{N}}}}}  = {q_0} + 2\sum\limits_{k = 1}^{K - 1} {{\rm{Re}}\left[ {{q_k}{e^{j2\pi \frac{{kn}}{{N}}}}} \right]},
\end{equation}
where the coefficients are conjugated symmetric as ${q_k} = q_{ - k}^*$, and $K$ is a number of non-zero coefficients in the one-sided frequency domain. Therefore, to design the waveform, we optimize the frequency coefficient vector ${\bf{q}}$ of size $K$ with $k$-th element ${q_k}$ $\left( {k = 0, \cdots ,K - 1} \right)$.

{\color{black}Before formulating a filter design problem, we need to update the existing average self-signal-to-noise-and-interference ratio (self-SINR) expression in \cite{kim2017waveform} using the stacked representation, and the self-SINR $\Upsilon \left( {\sigma_n^2 , {{\bf{q}}}} \right)$ can be defined as follows:
\begin{equation}\label{eq:self_SINR}
\Upsilon \left( {\sigma _n^2,{\bf{q}}} \right) \!
=\! \frac{1}{M}
\!\sum\limits_{i = 0}^{M - 1}
\! {\frac{{{{\left| {{{\left[ {{\bf{Q}}_f^H{{\bf{H}}_{{\rm{eff}}}}} \right]}_{\left( {i,i + N} \right)}}} \right|}^2}}}{{\sum\limits_{
j = 0
\atop j \ne i + N
}^{2N - 1} {{{\left| {{{\left[ {{\bf{Q}}_f^H{{\bf{H}}_{{\rm{eff}}}}} \right]}_{\left( {i,j} \right)}}} \right|}^2}}
\!+\! {{\left[ {{\bf{Q}}_f^H{{\bf{Q}}_f}} \right]}_{\left( {i,i} \right)}}\sigma _n^2}}},
\end{equation}
where the receiver filter ${{\bf{Q}}_f}$ can be ${{\bf{Q}}_f} = {\bf{Q}}_f^{\left( {{\rm{MF}}} \right)} = {\bf{TG}}_f^{\left( 0 \right)}$ for matched filter, and ${{\bf{Q}}_f} = {\bf{\hat Q}}_f^{\left( {{\rm{MMSE}}} \right)}$ for the MMSE filter as shown in \eqref{eq:Qf_MMSE}.

With reference to \cite{kim2017waveform}, the optimization problem can be formulated with the spectral and time localization constraints as follows:
\begin{equation}\label{eq:designproblem}
\begin{aligned}
& \mathop {\text{maximize}}\limits_{{{\bf{q}}}}
& & \Upsilon \left( {\sigma_n^2 , {{\bf{q}}}} \right) \\
& \text{subject to}
& & {\sigma _t} < {\epsilon_t},
\,\,\, \left| {q_0} + 2\sum\limits_{k = 1}^{K - 1} {{\rm{Re}}\left[ {{q_k}} \right]} \right| < {\epsilon_0}, \\
& & & \left| \sum\limits_{k = 1}^{K - 1} {{k}{\rm{Im}}\left[ {{q_k}} \right]} \right| < {\epsilon_1},
\,\,\,\left| \sum\limits_{k = 1}^{K - 1} {{k^2}{\rm{Re}}\left[ {{q_k}} \right]} \right| < {\epsilon_2},
\end{aligned}
\end{equation}
where ${\sigma _t}$ is a time dispersion parameter defined as
\begin{equation}
{\sigma _t}
= \frac{1}{N}
\sqrt {\sum\limits_{n = 0}^{N - 1} {{{\left( {n - \sum\limits_{n = 0}^{N - 1} {n{{\left| {p_0\left[ n \right]} \right|}^2}} } \right)}^2}{{\left| {p_0\left[ n \right]} \right|}^2}} }.
\end{equation}}
 \color{black}

\color{black} Since the prototype filter problem is non-convex with the high-dimensional variables, thus the filter design may require a high computational complexity. After the prototype filter is successfully designed, the system can simply utilize the designed filter at no extra design cost, however it is very difficult to redesign this filter every time the channel changes. Therefore, this paper assumes that the prototype filter is pre-designed before the transmission.
On the other hand, in equation \eqref{eq:self_SINR}, the self-SINR depends on the channel ${\bf{H}}$ and the noise variance of the receiver $\sigma _n^2$.
Since it is difficult to consider an actual channel value or a particular channel model when pre-designing the prototype filter, we assume that the channel is AWGN with ${{\bf{H}}_{{\rm{eff}}}} = {\bf{T}}{\overline {\bf{G}} _f}$.
About the the noise variance of the receiver $\sigma _n^2$, we set the target signal-to-noise ratio (SNR) to 15, 30, 50, and infinite dB, and design the prototype filter considering the MMSE receiver filter for each target SNR.

Since the optimization problem \eqref{eq:designproblem} cannot be solved with convex optimization tool, we design the prototype filters with $K = 15$ frequency domain filter taps and $L = 4$ oversampling factor through the pattern search algorithm which is a global optimization technique \cite{audet2004pattern}. The pattern search algorithm basically performs a polling process, which updates the optimal point by adding or subtracting each vector element from a given vector point. We set the self-SINR function $\Upsilon \left( {\sigma _n^2,{\bf{q}}} \right)$ for the frequency coefficient vector $\bf{q}$ as the fitness function, the fall-off rate conditions as the linear constraint, and the time dispersion condition as the nonlinear constraint. The tolerance parameters of each constraint are set to $\epsilon_0=0.01$ and $\epsilon_1=\epsilon_2=\epsilon_t=0.1$. \color{black}

\begin{figure}[!t]
\centering
\includegraphics[draft=false,width=0.99\columnwidth]{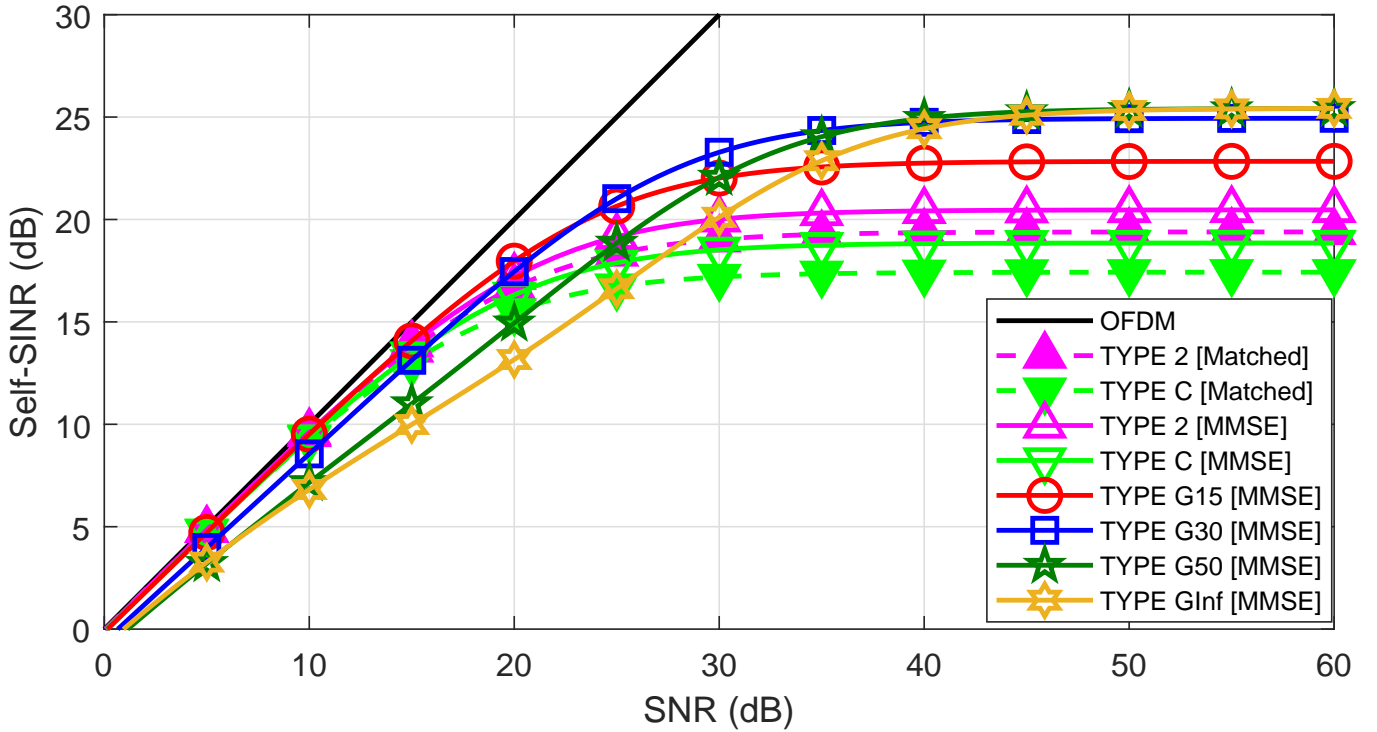}
{\caption{\color{black}{The self-SINR comparisons of the designed prototype filters with SNR variation.\color{black}}\label{fig_selfsinr}}}
\vspace{-3mm}
\end{figure}

The design results are described in Table \ref{tbl_filters}, and summarized in Table \ref{tbl_compare_filters}. The Fig. \ref{fig_selfsinr} shows the self-SINR performances of the each prototype filter with SNR variation. As intended in the filter design, the Type G15 and G30 filters exhibit the highest self-SINR at 15dB and 30dB SNR, respectively. Overall, the proposed prototype filters for the MMSE receiver filter show better performances than the reference filters for the matched receiver filter.
\color{black} Specifically, comparing the reference filter Type 2 and Type G30 at the SNR of 30dB, the self-SINR can achieve the performance improvement of 4.25dB, and in the case of 50dB SNR, the Type G50 improves the self-SINR of 5.98dB. \color{black}

\begin{table}[!t]
\setlength{\tabcolsep}{7pt}
\renewcommand{\arraystretch}{1.3}
\caption{Proposed filter coefficients in the frequency domain}\label{tbl_filters}
\begin{center}
\vspace{-2mm}
\begin{tabular}{|c||c|c|c|c|}
\hline
\multicolumn{1}{|c||}{\multirow{2}{*}{}} & \multicolumn{1}{c|}{Type G15} & \multicolumn{1}{c|}{Type G30} & \multicolumn{1}{c|}{Type G50} & \multicolumn{1}{c|}{Type Ginf}\\ \cline{2-5}
\multicolumn{1}{|c||}{}                  & \multicolumn{4}{c|}{Only real coefficients}                                                \\
\hline\hline
\vspace{-1mm}$q_0$     &  +1.0000    & +1.0000   &  +1.0000   &  +1.0000  \\
\vspace{-1mm}$q_1$     &  -0.8660    & -0.9591   &  -0.9988   &  -0.9655  \\
\vspace{-1mm}$q_2$     &  +0.6662    & +0.7533   &  +0.7628   &  +0.7616  \\
\vspace{-1mm}$q_3$     &  -0.3932    & -0.3915   &  -0.3597   &  -0.4422  \\
\vspace{-1mm}$q_4$     &  +0.0066    & +0.0844   &  +0.1029   &  +0.1859  \\
\vspace{-1mm}$q_5$     &  +0.2122    & +0.2388   &  +0.1849   &  +0.0610  \\
\vspace{-1mm}$q_6$     &  -0.2680    & -0.4369   &  -0.3509   &  -0.1987  \\
\vspace{-1mm}$q_7$     &  +0.1702    & +0.2274   &  +0.2325   &  +0.1965  \\
\vspace{-1mm}$q_8$     &  -0.0050    & -0.0648   &  -0.1383   &  -0.1652  \\
\vspace{-1mm}$q_9$     &  -0.1571    & -0.0774   &  -0.0504   &  -0.0025  \\
\vspace{-1mm}$q_{10}$  &  +0.1871    & +0.2961   &  +0.3059   &  +0.1745  \\
\vspace{-1mm}$q_{11}$  &  -0.0910    & -0.2442   &  -0.2552   &  -0.1267  \\
\vspace{-1mm}$q_{12}$  &  +0.0091    & +0.1152   &  +0.1836   &  +0.1106  \\
\vspace{-1mm}$q_{13}$  &  +0.1623    & -0.0077   &  -0.2207   &  -0.1760  \\
             $q_{14}$  &  -0.1315    & -0.0317   &  +0.1033   &  +0.0889  \\
\hline
\end{tabular}
\end{center}
\vspace{-3mm}
\end{table}

\begin{table}[!t]
\setlength{\tabcolsep}{4pt}
\renewcommand{\arraystretch}{1.6}
\caption{Comparison of the prototype filters\label{tbl_compare_filters}}
\begin{center}
\vspace{-2mm}
\begin{tabular}{c||c|c|ccccccccccccc}
\hline\hline

& \multicolumn{1}{c|}{Ref. \cite{yun2015new}}
& \multicolumn{1}{c|}{Ref. \cite{kim2017waveform}}
& \multicolumn{4}{c}{Proposed}\\

\cline{1-7}
Type  & 2  & C  & G15 & G30 & G50 & Ginf\\
\hline\hline
 Receiver filter   & \multicolumn{2}{c|}{Matched} & \multicolumn{4}{c}{MMSE} \\
 \cline{2-7}
 Target SNR (dB)  & $\infty$ & $\infty$ & 15 & 30 & 50 & $\infty$\\
 $K$(taps)   & 15 & 15 & 15 & 15 & 15 & 15\\
 $\Upsilon \left( {0 ,{\bf{q}}} \right)$ (dB) & 19.2 & 17.4 & 22.8 & 24.9 & 25.4 & 25.4\\
 Fall-off rate & ${\left| \omega  \right|^{ - 4}}$ & ${\left| \omega  \right|^{ - 5}}$ & ${\left| \omega  \right|^{ - 5}}$ & ${\left| \omega  \right|^{ - 5}}$ & ${\left| \omega  \right|^{ - 5}}$ & ${\left| \omega  \right|^{ - 5}}$\\
 ${\sigma _t}$ & 0.197 & 0.083 & 0.077 & 0.078 & 0.075 & 0.064\\
 Coefficient & Complex & Real & Real & Real & Real & Real\\
\hline\hline
\end{tabular}

\end{center}
\vspace{-3mm}
\end{table}

\color{black}\section{Simulation Results}\label{sec:simul}

In this section, we compare the performances of the designed prototype filters, which are presented in Table \ref{tbl_compare_filters}. \color{black}For reference filters with matched receiver filter, we use a single-tap MMSE equalizer in the upsampled frequency domain. Also, we consider the performances of the reference filters with MMSE receiver filter, which is same as the process of the proposed prototype filter. \color{black}

\color{black} The common simulation parameters are $L=4$, $M=128$, $\sigma _d^2 = 1$, 16-QAM/64-QAM, 2GHz carrier frequency, and 15kHz subcarrier spacing.
The channel models of the simulation are AWGN, extended ITU pedestrian A (EPA), and extended ITU vehicular A (EVA) \cite{3gpp2008evolved}. \color{black}

\begin{figure}[p]
\centering
   \subfigure{
     	\includegraphics[draft=false,width=0.99\columnwidth] {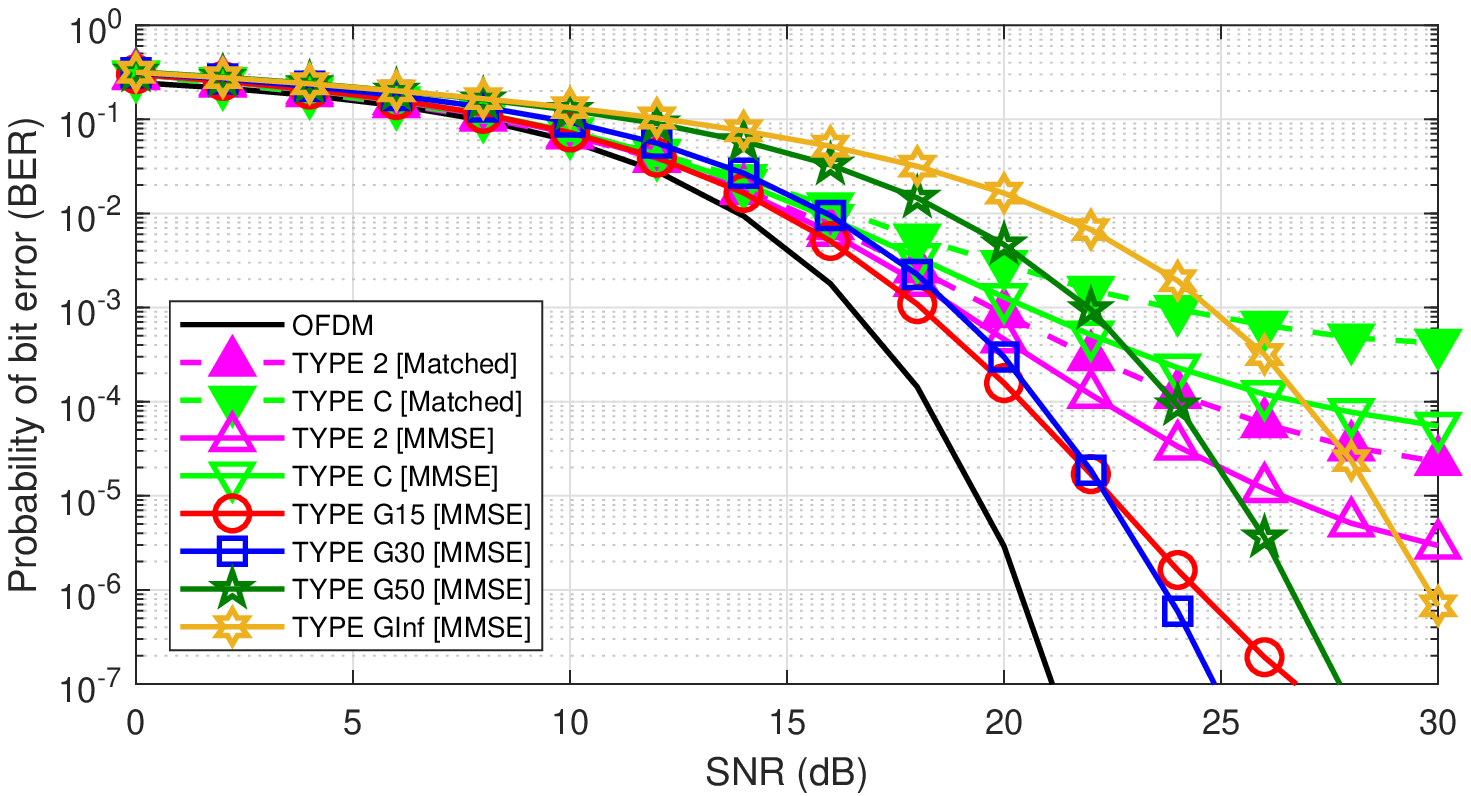}
	}
	\subfigure{
     	\includegraphics[draft=false,width=0.99\columnwidth] {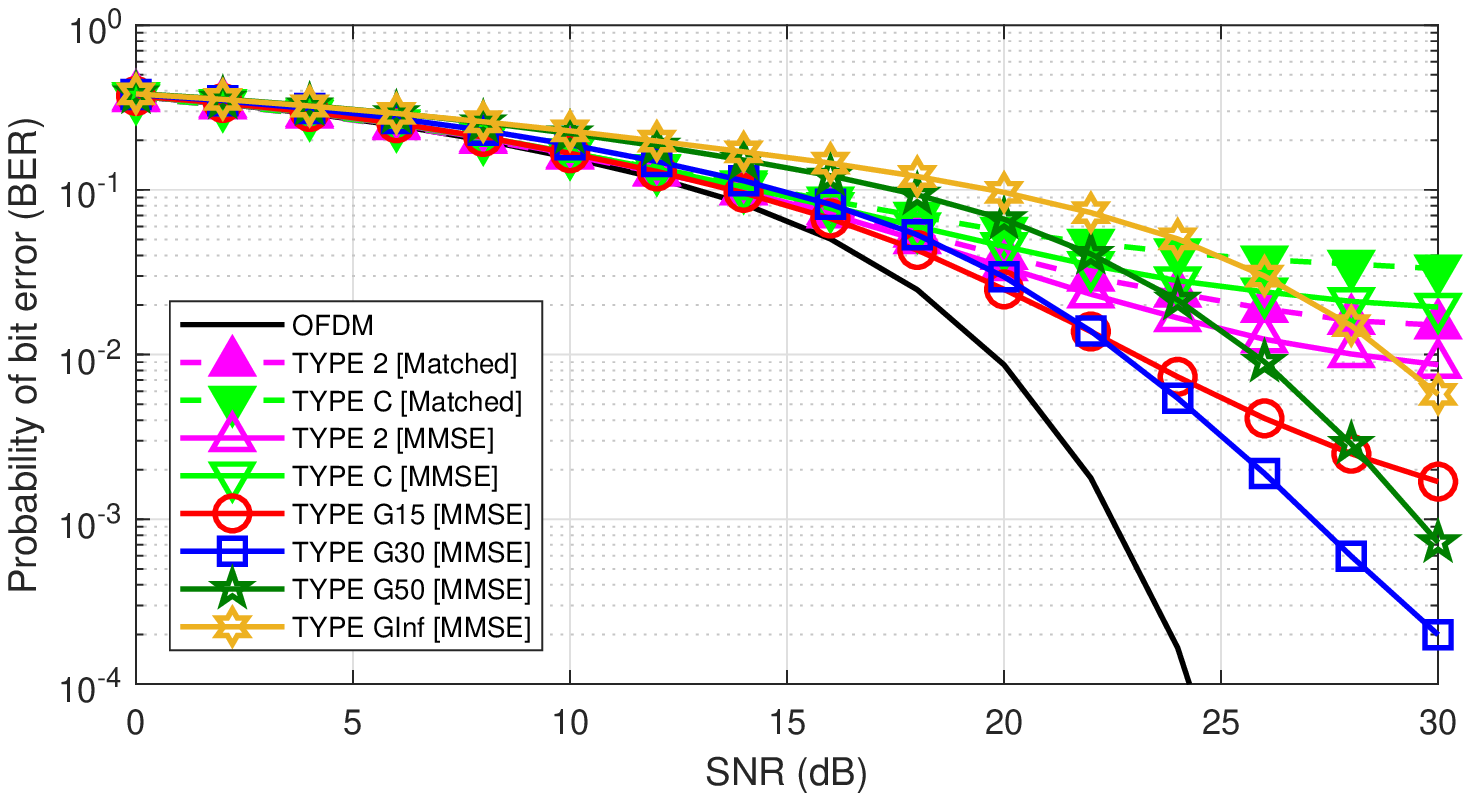}
	}
{\caption{\color{black}The BER performance comparisons in the AWGN channel with 16-QAM (top) and 64-QAM (bottom).\color{black}}\label{fig:BER_awgn}}
\vspace{-3mm}
\end{figure}

\begin{figure}[p]
\centering
   \subfigure{
     	\includegraphics[draft=false,width=0.99\columnwidth] {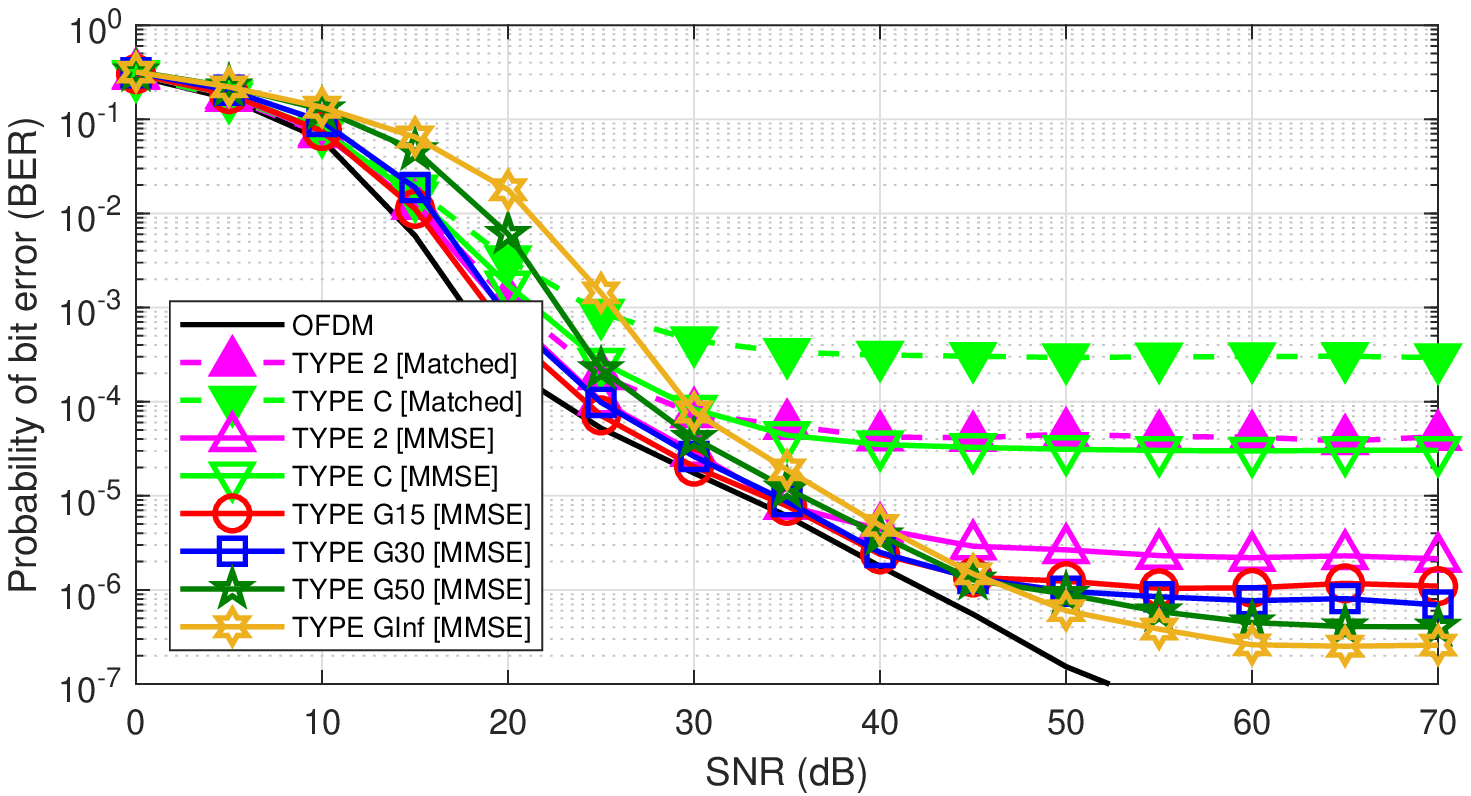}
	}
	\subfigure{
     	\includegraphics[draft=false,width=0.99\columnwidth] {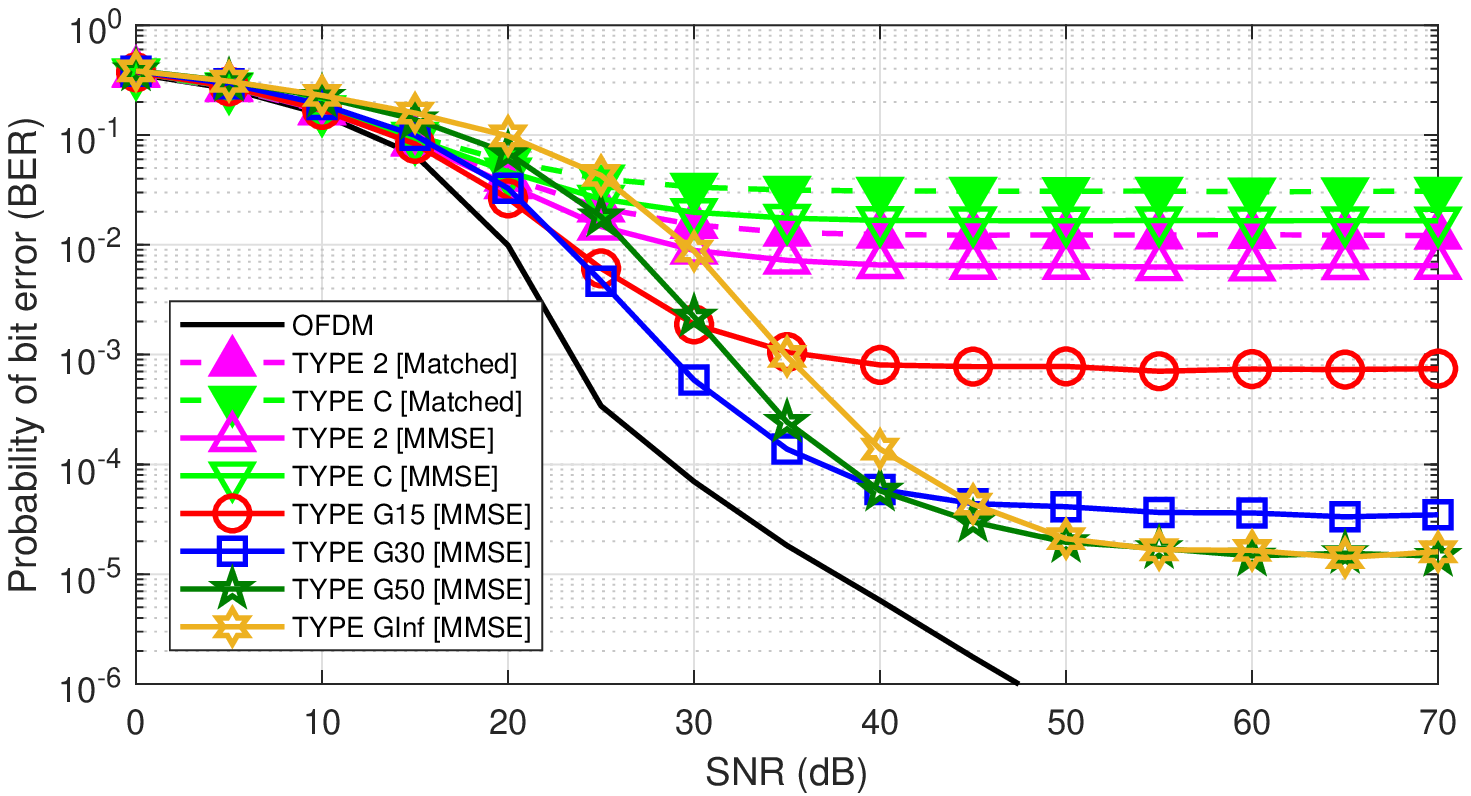}
	}
{\caption{\color{black}The BER performance comparisons in the EPA channel with 16-QAM (top) and 64-QAM (bottom).\color{black}}\label{fig:BER_pede}}
\vspace{-2mm}
\end{figure}

\begin{figure}[p]
\centering
   \subfigure{
     	\includegraphics[draft=false,width=0.99\columnwidth] {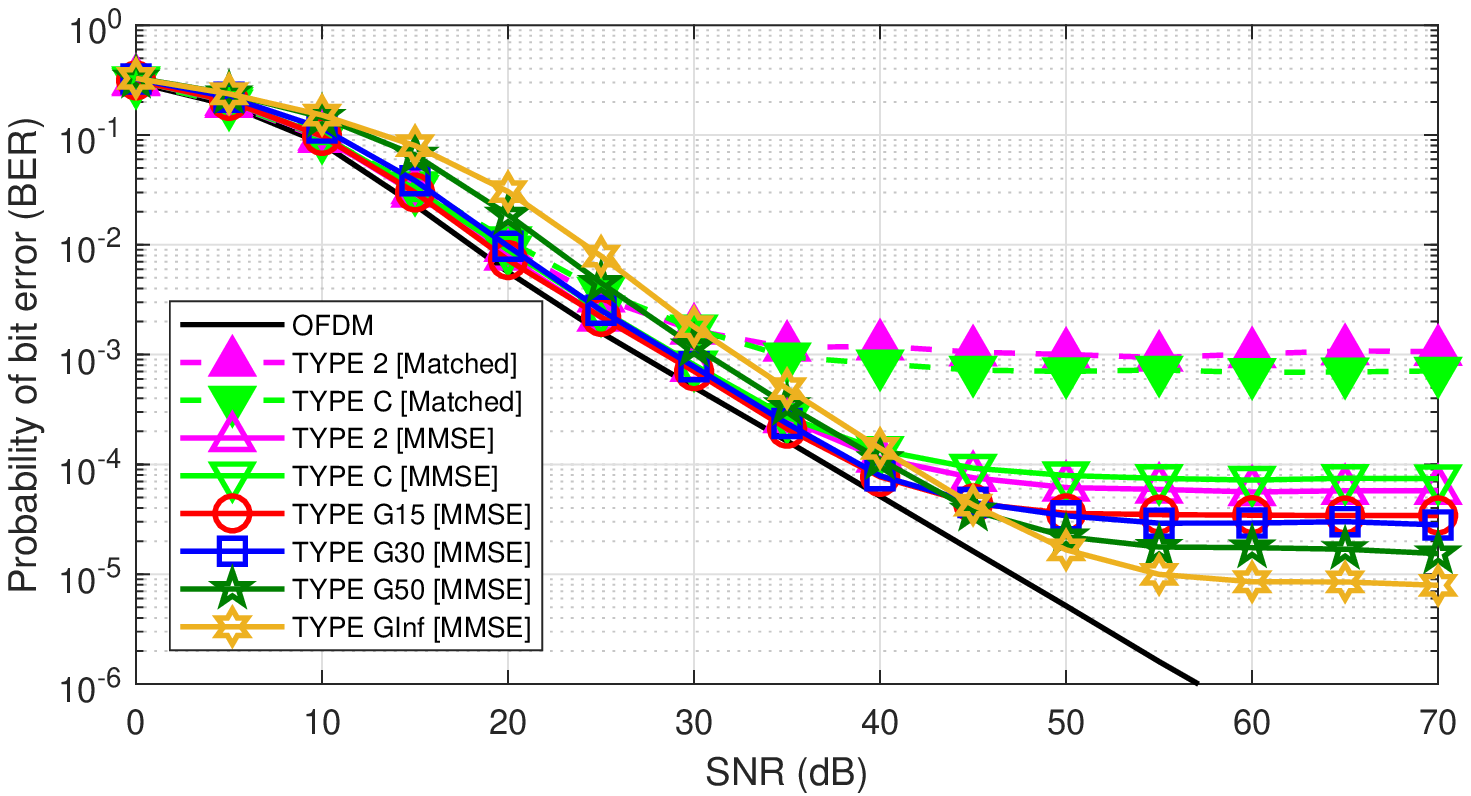}
	}
	\subfigure{
     	\includegraphics[draft=false,width=0.99\columnwidth] {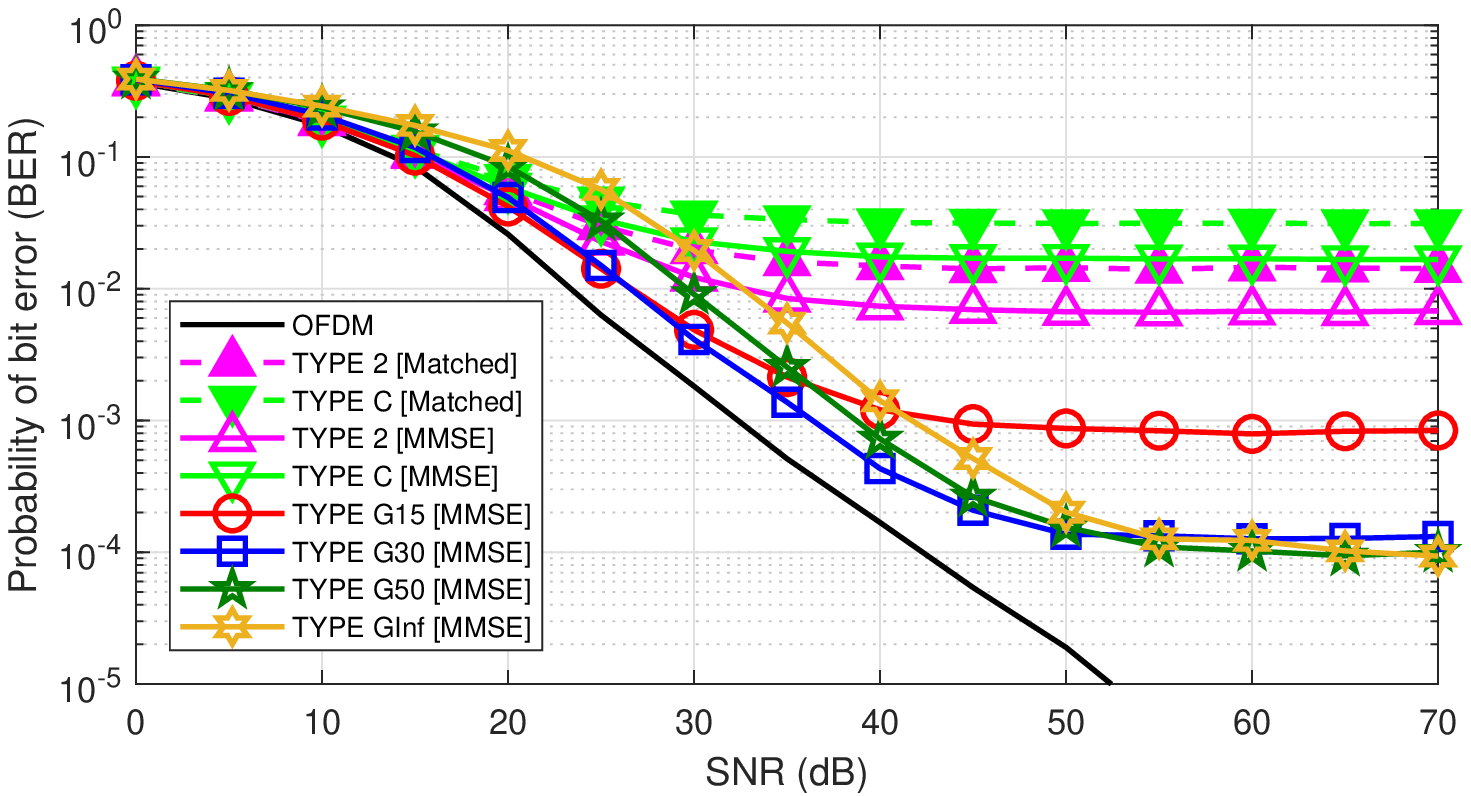}
	}
{\caption{\color{black}The BER performance comparisons in the EVA channel with 16-QAM (top) and 64-QAM (bottom).\color{black}}\label{fig:BER_vehi}}
\vspace{-2mm}
\end{figure}

In Fig. \ref{fig:BER_awgn}, \ref{fig:BER_pede}, and \ref{fig:BER_vehi}, we compare the probability of bit error (BER) performances between the reference filters with matched receiver filter and the proposed filters with the MMSE receiver filter.
In the simulation, as mentioned above, we can see that the prototype filters with higher self-SINR show better BER performances as expected, and the performances of the proposed filters are significantly enhanced than the reference filters. As shown in Fig. \ref{fig:BER_awgn}, the Type G30 filter with the highest self-SINR at 30dB SNR shows the best performance in AWGN. And in EPA and EVA channels which require higher SNR, Type G50 and GInf show slightly better BER performances than Type G30. Overall, we can expect that the Type G30 filter performs well in most of the scenarios, so we expect it to be usable in general situation. \color{black} Additionally, we check the BER performances of the reference filters with MMSE receiver filter. Even if the signal with the reference filters is received through MMSE filter, the reference filters show the slightly better performances than the matched receiver, but still worse than the proposed filter because it is not designed to use the MMSE filter. \color{black}

\color{black} The most important part in these BER performance results is that the performance degradation of the reference filter compared to OFDM can be significantly mitigated through the proposed prototype filter and MMSE receive structure.
Specifically, by the proposed filter and receiver, we can reduce the residual interference so that QAM-FBMC can achieve $10^{-5}$ uncoded BER performance in high modulation order. In particular, this MMSE receiver works in the linear system instead of complex nonlinear operations such as successive cancellation, and it can be realizable when QAM-FBMC is used in a realistic environment. \color{black}

\color{black} Additionally, we can confirm that the designed prototype filters assuming AWGN channel show almost the same performance trends as for AWGN, even if BER simulations are performed on the fading channel models.
In the practical system, if the QAM-FBMC transmitter knows or predicts the received SNR, the system can use the pre-designed prototype filters corresponding to the target SNR, which may exhibit the best BER performance.
Therefore, if the QAM-FBMC system utilizes the pre-designed prototype filters appropriately, we can expect the good performance even considering the practical channel and varying SNR at the receiver. \color{black}

\begin{figure}[!t]
\centering
\includegraphics[draft=false,width=0.99\columnwidth]{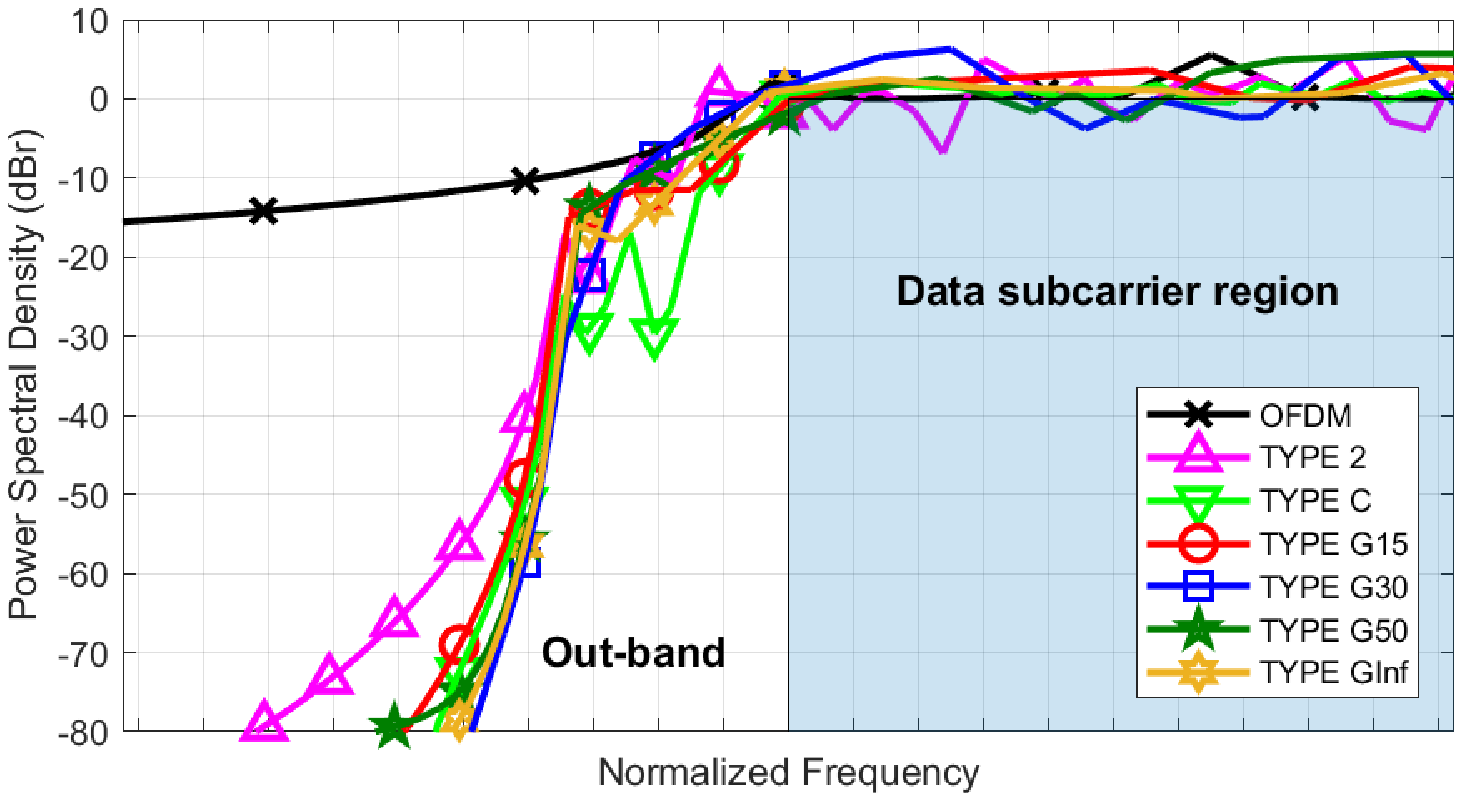}
{\caption{\color{black}The comparison of power spectral density between OFDM and various QAM-FBMC prototype filters.\color{black}\label{fig:REV_fig_PSD}}}
\vspace{-2mm}
\end{figure}

\color{black} Fig. \ref{fig:REV_fig_PSD} shows the power spectral density (PSD) of the prototype filters for comparing the spectral confinement characteristics. As we can expect from Table \ref{tbl_compare_filters}, the Type C and the proposed filters all show the almost same PSD, with a good spectral confinement of ${\left| \omega  \right|^{ - 5}}$, the Type 2 has a medium spectral confinement with a fall-off rate of ${\left| \omega  \right|^{ - 4}}$, and OFDM shows the worst characteristic.

{\color{black}\section{Conclusions}\label{sec:conclusion}

In this paper, we designed waveforms for QAM-FBMC considering an MMSE receiver filter.} {\color{black}In a QAM-FBMC system \color{black}using a non-orthogonal pulse-shaping filter, the conventional matched receiver filter cannot be optimal, and we proposed a MMSE filter to minimize interference.} {\color{black}In that regard, we \color{black}simplified the linear system model for QAM-FBMC using stacked vectorization, and formulated a receiver filter following the MMSE criterion.}
\color{black} In addition, we designed prototype filters that had a superior self-SINR under the condition of the linear receiver. Finally, we confirmed that the proposed filters show the best performance with the design intent through simulation results.
\color{black} In addition, the formulations of the linear system and self-SINR equations for QAM-FBMC showed the simpler representation than that of the previously complex QAM-FBMC system, therefore, we identified the possibility of applying popular signal processing technology directly to QAM-FBMC in the future. \color{black}

\clearpage


%


%


\ifCLASSOPTIONcaptionsoff
  \newpage
\fi

\bibliographystyle{./IEEEtran}
\bibliography{./IEEEabrv,./Han_bib}

\end{document}